\documentclass[10pt]{article}
\usepackage{graphics}

\usepackage{graphicx}
\input{psfig.sty}

\newcommand{\bs}{\bf}
\newcommand{\boldsymbol}{\bs}
\newcommand{\nn}{\nonumber}
\newcommand{\FP}{\mathcal J}
\newcommand{\be}{\begin{equation}}
\newcommand{\ee}{\end{equation}}
\newcommand{\ar}{\arrowvert}

\newcommand{\ba}{\begin{eqnarray}}
\newcommand{\ea}{\end{eqnarray}}

\begin{document}

\noindent
\begin{center}
\ \center\small{COULOMB GAUGE MODEL FOR EXOTIC HADRONS}
\ \center\small{STEPHEN R. COTANCH$^a$ and FELIPE LLANES-ESTRADA$^b$}
\ \center{$^a$\em{Department of Physics, North Carolina State University, Raleigh, 
NC 27695, USA}}
\ \center{$^b$\em{Depto. F\'isica Te\'orica I, Universidad Complutense de Madrid,
28040 Madrid, Spain}}
\end{center}


\noindent
Hadron structure advances provided by the Coulomb gauge model are  summarized.
Highlights include a realistic description of the vacuum and meson spectrum with chiral
symmetry and dynamical flavor mixing,  accurate hyperfine splittings for light and heavy systems, glueball and hybrid meson predictions in agreement with lattice QCD, and tetraquark results providing an understanding of the observed $\pi_1(1400)$ having unconventional quantum numbers.
\\
\hspace{-1.5em}PAC numbers: 11.30.Rd, 12.38.Lg, 12.39.Mk, 12.40.Vv, 12.40.Yx
\\
\hspace{-1.5em}Keywords:  Coulomb gauge Hamiltonian,  glueballs, hybrid mesons, tetraquarks  

\begin{center}
\center{\Large\em{1.  Introduction}}
\\
\end{center}

Understanding whether exotic systems do or do not exist is one of the few remaining challenges for the Standard Model.
Such states are those with explicit gluon, $g$,  degrees of freedom and/or quarks, $q$, in color states that are not singlets. Hadrons with unconventional quantum numbers, i.e. those not possible in $q \bar q$ or $qqq$ systems such as $J^{PC} = 1^{-+}$, may or may not be exotic. For example, one of the key predictions of the Coulomb gauge model described in this paper is that the observed $1^{-+}$ $\pi_1(1400)$ is not an exotic but rather a more conventional meson-meson resonance with   quarks in two color singlet states. Addressing the above challenge has motivated the development and application of this model which  provides a robust, relativistic field theoretical approach that  retains the attractive, insightful wavefunction picture. Since QCD is  intrinsically a non-perturbative many-body problem, an established, successful nuclear structure method has been adopted which entails three elements: 1) an effective Hamiltonian; 2) a truncated model space; 3) solving the equations of motion using standard many-body techniques. The Coulomb gauge model therefore uses several many-body techniques, Bardeen-Cooper-Schrieffer [BCS], Tamm-Dancoff [TDA],
random phase approximation [RPA], coupled channels, variational and exact diagonalization to solve the equations of motion for an approximate QCD Hamiltonian in a model space spanned by a truncated number of particle Fock states. For this method to work the Fock space expansion,  $|\Psi> = |q\bar q> + |gg> + |q \bar q g>
+ |q\bar qq\bar q> + ...$, must involve dressed (constituent) quaisparticle states, and not  bare (current) quarks and gluons.  This is because  the dressed partons have much larger masses which, from energy considerations,  enables a sensible truncation to describe the low mass hadron spectrum in analogy to the reasonable 1 particle-1 hole, 2 particle-2 hole, etc. truncations  for  low lying nuclear states.

The following sections detail the Coulomb gauge model beginning with the Hamiltonian in section 2 and the BCS vacuum treatment for dressing the bare quarks and gluons in section 3.  Applications to hadrons are discussed in section 4 for two-body $q \bar q$ (mesons) and $gg$ (glueballs) states,
section 5 for three-body $ggg$ (oddballs) and $q \bar q g$ (hybrids) states,  and section 6 for four-body $q \bar qq \bar q$ (tetraquarks) states. Section 6 also addresses dynamic mixing of two and four-body states.  Finally, a summary is presented in section 7.
\\

\begin{center}
\center{\Large\em{2.   QCD Coulomb gauge model}}
\\
\end{center}

In the Coulomb gauge the exact QCD Hamiltonian 
 is
\begin{eqnarray}
H_{\rm QCD} &=& H_q + H_g +H_{qg} + H_{C}    \\
H_q &=& \int d{\bs x} \Psi^\dagger ({\bs x}) [ -i
{\mbox{\boldmath$\alpha$\unboldmath}} \cdot
{\mbox{\boldmath$\nabla$\unboldmath}}
+  \beta m] \Psi ({\bs x})   \\
H_g &=& \frac{1}{2} \int\!\! d {\bs x} \; \!\! \left[ \FP^{-1}{\bf
\Pi}^a({\bs x})\cdot \!\!  \FP {\bf
\Pi}^a({\bs x}) +{\bf B}^a({\bs x})\cdot{\bf B}^a({\bs x}) \right] \; \;  \; \; \\
H_{qg} &=&  g \int d {\bs x} \; {\bf J}^a ({\bs x})
\cdot {\bf A}^a({\bs x}) \label{eq:J.A}\\
H_C &=& -\frac{g^2}{2} \int d{\bs x} d{\bs y}\FP^{-1} \rho^a ({\bs x})
 K^{ab}( {\bs x},{\bs y}  ) \FP \rho^b ({\bs y})   \ ,
\end{eqnarray}
where $g$ is the QCD coupling constant, $\Psi$ is the quark field with current
quark mass $m$, ${A}^a =({\bf A}^a, A^a_0)$ are the gluon fields satisfying the
transverse gauge condition, ${\bs{\nabla}\cdot}{\bf A}^a = 0$ $(a = 1, 2, ... 8)$,
 ${\bf \Pi}^a = -{ \bf E}^a_{tr} $ are the conjugate momenta,
  \begin{eqnarray}
\mathbf{E}_{tr}^a&=&-\dot{\mathbf{A}}^a + g( 1- {\mbox{$\nabla$\unboldmath}}^{-2}
{\mbox{\boldmath$\nabla$\unboldmath}}{\mbox{\boldmath$\nabla$\unboldmath}} \cdot  )f^{abc} A^{b}_0\mathbf{A}^c  \\
{\bf B}^a &=& \nabla \times {\bf A}^a + \frac{1}{2} g f^{abc} {\bf
A}^b \times {\bf A}^c \ ,
\end{eqnarray}
are the non-abelian chromodynamic fields,
${\bf J}^a = \Psi^\dagger ({\bs x})
\mbox{\boldmath$\alpha$\unboldmath}T^a \Psi ({\bs x})$ are the quark currents,
$\rho^a({\bs x}) = \Psi^\dagger({\bs x}) T^a\Psi({\bs x})
+f^{abc}{\bf
A}^b({\bs x})\cdot{\bf \Pi}^c({\bs x})$
are the color densities
with standard    $SU(3)$
color matrices,  $T^a = \frac{\lambda^a}{2}$, and structure constants, $f^{abc}$.
The gauge manifold curvature is measured by
the
Faddeev-Popov determinant, $\FP = {\rm det}(\mathcal M)$, of the
matrix, ${\mathcal M} = {\mbox{\boldmath$\nabla$\unboldmath}} \cdot
{\bf D}$, with covariant derivative, ${\bf D}^{ab} =
\delta^{ab}{\mbox{\boldmath$\nabla$\unboldmath}}  - g f^{abc} {\bf
A}^c$, which governs
the kernel, $K^{ab}({\bs x}, {\bs y}) = \langle{\bs x},
a|{\mathcal M}^{-1} \nabla^2 {\mathcal M}^{-1}  |{\bs y}, b\rangle$.
In this gauge, the color form of Gauss's law, which
is essential for confinement, is satisfied exactly and can be used to
eliminate the unphysical
longitudinal  gluon fields.
The Hamiltonian is renormalizable, preserves rotational invariance, avoids
spurious retardation corrections, aids identification of dominant,
low energy  potentials, permits resolution
of the Gribov problem and  has only physical degrees of
freedom (no ghosts).
The standard  normal mode field expansions are
(bare quark spinors $u, v$, helicity, $\lambda = \pm 1$, and color
vectors $\hat{{\epsilon}}_{{\cal C }= 1,2,3}$)
\begin{eqnarray}
\label{colorfields1}
 \Psi(\boldsymbol{x}) &=&\int \!\! \frac{d
   {\bs{k}}}{(2\pi)^3} [{u}_{\lambda}
({\boldsymbol{k}}) b_{\lambda {\cal C}}({\boldsymbol{k)}}  +
{v}_{\lambda} (-{\boldsymbol{k}})
    d^\dag_{\lambda {\cal C}}({\boldsymbol{-k)}} ] e^{i \boldsymbol{k} \cdot \boldsymbol{x}} \hat{{\epsilon}}_{\cal C}  \\
{\bf A}^a({\bs{x}}) &=&  \int \frac{d{\bs{k}}}{(2\pi)^3}
\frac{1}{\sqrt{2k}}[{\bf a}^a({\bs{k}}) + {\bf
a}^{a\dag}(-{\bs{k}})] e^{i{\bs{k}}\cdot {\bs {x}}}  \ \ \
\\
{\bf \Pi}^a({\bs{x}}) &=& \hspace{-.15cm}-i \int
\!\!\frac{d{\bs{k}}}{(2\pi)^3} \sqrt{\frac{k}{2}} [{\bf
a}^a({\bs{k}})-{\bf a}^{a\dag}(-{\bs{k}})]e^{i{\bs{k}}\cdot
{\bs{x}}}  \!.
\end{eqnarray}
Here
$b_{\lambda {\cal C}}(\boldsymbol{k)}$, $d_{\lambda {\cal
C}}(\boldsymbol{-k)} $ and  $a_{\mu}^a({\bs{k}})$ ($\mu = 0, \pm 1$)
are the respective  quark, anti-quark and gluon Fock operators, the latter
satisfying   the transverse commutation relations
\begin{eqnarray}
[a^a_{\mu}({\bs k}),a^{b \dagger}_{\mu'}({\bs k}')] &=& (2\pi)^3
\delta_{ab} \delta^3({\bs k}-{\bs k}')[\delta_{{\mu}{\mu'}}-
(-1)^{\mu}\frac{k_{\mu} k_{-\mu'}}{k^2}] \nonumber \\
&=& (2\pi)^3
\delta_{ab} \delta^3({\bs k}-{\bs k}') D_{\mu \mu'}({\bf k}) \ ,
\end{eqnarray}
due to the Coulomb gauge  condition, ${\bf k}\cdot {\bf a}^a ({\bf k}) = (-1)^\mu k_{\mu} a_{-\mu} ^a ({\bs k}) =0$. 

The Coulomb gauge model Hamiltonian, $H_{CG}$, is obtained by replacing the Coulomb kernel with a calculable confining potential and
using the lowest order, unit value for the the Faddeev-Popov determinant
  \begin{eqnarray}
H_{\rm CG} &=& H_q + H_g^{\rm CG} + H_{qg}^{\rm CG} + H_C^{\rm CG}   \\
H_g^{\rm CG} &=& \frac{1}{2} \int d {\bs x}\left[ {\bf
\Pi}^a({\bs x})\cdot {\bf \Pi}^a({\bs x}) +{\bf B}^a({\bs
x})\cdot{\bf B}^a({\bs x})
\right] \label{eq:non-abelian} \\
H_{qg}^{\rm CG} &=&  \frac{1}{2} \int d{\bf x} d{\bf y}
J^a_{i} ({\bf x}) 
\left( \delta_{ij} - \frac{\nabla_i \nabla_j}{\nabla^2} \right)_{\bf x}
 {U}(|{\bf x}-{\bf y}|)J^a_{j} ({\bf y})   \\
H_C^{\rm CG} &=& -\frac{1}{2} \int d{\bs x} d{\bs y} \rho^a ({\bs
x}) {V}(\ar {\bs x}-{\bs y} \ar ) \rho^a ({\bs y})   \ .
 \label{model}
\end{eqnarray}
Also, using Maxwell's equation, the quark-gluon Hamiltonian has been converted to a quark-quark  hyperfine interaction involving an effective massive ($m_g =$ 600 MeV) gluon exchange potential (modified Yukawa kernel)
\begin{equation}
    { U}(p) = \left\{ \begin{array}{ll}
         -\frac{8.07}{p^2}
    \frac{\ln^{-0.62}(\frac{p^2}{m_g^2}+0.82)}{\ln^{0.8}(\frac{p^2}{m_g^2}+1.41)} & \textrm{$ p>m_g $} \\
        -\frac{5.509}{p^2+m_g^2} & \textrm{ $p<m_g$}    \nn 
        \end{array} \right\} \ .
\end{equation}
Confinement is provided by the
Cornell  potential, ${ V}(r)=-{\alpha_s}/{r}+\sigma r$, or
in momentum space,
$ {\ V}(p) = -4\pi {\alpha_s}/{p^2} - 8 \pi \sigma/ p^4$.
The string tension, $\sigma=0.135$ GeV$^{2}$, and
$\alpha_s =0.4$ have been independently determined
and set the scale for the calculation.  The remaining parameters are the bare (current) quark masses, $m_u = m_d = 5$ MeV, $m_s = 80$ MeV, $m_c = 640$ MeV and $m_b = 3.33$ GeV. 

\newpage

\begin{center}
\center{\Large\em{3.  BCS vacuum and  gap equations}}
\\
\end{center}

The Bardeen-Cooper-Schriffer  method is used to obtain the ground state (vacuum).  This entails  rotating the field operators
\begin{eqnarray} \label{eq:operator rotations}
    B_{\lambda {\cal C}}(\boldsymbol{k)} &=& \cos\frac{\theta_k}{2}
    b_{\lambda {\cal C}}({\boldsymbol{k})}  - \lambda \sin\frac{\theta_k}{2}
    d^\dag_{\lambda {\cal C}}(\boldsymbol{-k)}   \nonumber  \\ \nonumber
    D_{\lambda {\cal C}}(\boldsymbol{-k)}&=& \cos\frac{\theta_k}{2}
    d_{\lambda {\cal C}}({\boldsymbol{-k})}  + \lambda \sin\frac{\theta_k}{2}
    b^\dag_{\lambda {\cal C}}(\boldsymbol{k)}  \\
    { {\mbox{\boldmath$\alpha$\unboldmath}} }^a(\boldsymbol{k}) &=& \cosh \Theta_k {\bf a}^a({\boldsymbol{k}}) + \sinh \Theta_k
    {{\bf a}^a}^\dag(-\boldsymbol{k}) \ ,
\end{eqnarray}
producing the dressed,
quasi-particle operators ${ {\mbox{\boldmath$\alpha$\unboldmath}} }^a$, $B_{\lambda {\cal C}}$ and $D_{\lambda {\cal C}}$. 
The BCS quasi-particle vacuum, $|\Omega \rangle$, is defined by $B_{\lambda {\cal C}} |\Omega
\rangle = D_{\lambda {\cal C}} |\Omega\rangle = \alpha^a_\mu |\Omega\rangle = 0$,  and
builds on the bare parton vacuum,  $|0\rangle$, $b_{\lambda {\cal C}} |0\rangle = d_{\lambda {\cal C}} |0\rangle = a^a_\mu
|0\rangle = 0$,
\begin{equation}
    |\Omega_{quark}\rangle =[e^{- \int \!\!
    \frac{d\boldsymbol{k}}{(2\pi)^3}\lambda
    \tan\frac{\theta_k}{2} b^\dag_{\lambda {\cal C}}({\boldsymbol{k}})
    d^\dag_{\lambda {\cal C}} (-\boldsymbol{k})}] \;  |0\rangle   \nonumber
\end{equation}
\begin{equation}
    |\Omega_{gluon}\rangle = [e^ {- \!\! \int \!\!
    \frac{d\boldsymbol{k} }{(2\pi)^3} \frac{1}{2}\tanh\Theta_k D_{\mu \mu'}({\bs k})
    a_{\mu}^{a\dag} ({\boldsymbol{k}}) a_{\mu'}^{a\dag} (-\boldsymbol{k})}] \; |0\rangle \ . \nonumber
\end{equation}
Quark and gluon condensates (correlated $q\bar{q}$ and $gg$ Cooper pairs) naturally emerge in the composite BCS vacuum, $|\Omega \rangle = |\Omega_{quark}
\rangle \otimes |\Omega_{gluon} \rangle $.
 A variational minimization of the model ground state, $\delta \langle\Omega|H_{\rm CG}|
\Omega\rangle = 0$,
yields the constituent quark and gluon  gap equations
\begin{eqnarray}
k \ s_k - m_q \ c_k &=& \frac{2} {3}\int \frac{d{\bf q}}{(2\pi)^3}
\left[
V(\ar {\bf k} - {\bf q} \ar) (s_k c_q x - s_q c_k )
\right.  \nonumber \\ 
&&\left. -4 (c_k s_q U(\ar {\bf k} - {\bf q}\ar)  - c_q s_k W(\ar {\bf k} - {\bf
q}\ar) ) \right]  \\
 W({|\bf k} - {\bf q}|) &\equiv& U(|{\bf k} - {\bf q}|)
\frac{x(k^2+q^2)-qk(1+x^2)}{\ar{\bf k}-{\bf q}\ar^2} 
\\
\label{ggapeq}
    \omega_k^2 = k^2- \frac{3}{4} 
    \int  &&\!\! \! \! \!\! \! \! \!\! \! \! \!\! \! \!
    \frac{d\boldsymbol{q}}{(2\pi)^3}
[V(|{\boldsymbol{k-q}}|)
     \frac{(1 + x^2)(\omega_q^2-\omega_k^2)}{\omega_q}  -g^2  \frac{(1 - x^2)}{\omega_q}] \ , 
 \ea
with $s_k = sin \phi_k$, $c_k = cos \phi_k$ and $x = {\bs k
}\cdot {\bs q}$.
Here $\phi_k = \phi (k)$ is  the quark gap angle
related to the
BCS angle $\theta_k$ by, $tan(\phi_k  - \theta_k ) = m/k$, and
$\omega_k = k e^{-2\Theta_k}$ is the effective  gluon self energy.
The quark gap
equation is UV finite for the linear potential, 
$ - 8 \pi \sigma /p^{4}$, but
not for the Coulomb potential,
$ - 4 \pi \alpha_s /p^{2}$.   The gluon gap equation has both logarithmical and quadratical UV
divergences  and an integration cutoff, $\Lambda
= 4$ GeV, determined in previous studies is used in both equations.
The gap equations yield the  quark, $E_k = \sqrt{M(k)^2 + k^2} = M(k)/sin \phi_k$,
 and  gluon,
 $\omega_k$,   self-energies from which the dressed (constituent) quasi-particle masses can
 be extracted at zero momentum: $M_u (0) \cong 125 $ MeV,  $M_g (0) = \omega(0)  \cong 800 $ MeV.
The gap angles also determine the quark,
${\langle q \bar {q} \rangle} = \langle \Omega | \bar {\Psi}(0) \Psi(0)| \Omega \rangle
  =- (177 \, \rm MeV)^3 $,
  and gluon, 
 $ \langle \alpha G^a_{\mu \nu} G_a^{\mu \nu} \rangle
  = (433 \, \rm  MeV)^4 $, condensates
 which are in reasonable agreement  with   QCD sum  rule,  $ -(236 \, \rm MeV)^3$, and lattice, $(441 \, \rm MeV)^4$, values, respectively.
\\

\begin{center}
\center{\Large\em{4.  Meson and glueball states}}
\\
\end{center}

Turning to excited states, the lightest in the quark and gluon sectors are two-body mesons ($q\bar q$) and glueballs ($gg$). The meson states were calculated using both the TDA and RPA, the latter
being essential to preserve chiral symmetry.  The TDA and RPA states
are respectively,  $\arrowvert \Psi^{nJPC}_{TDA} \rangle =
Q^{\dagger}_{nJPC}(TDA)\arrowvert
\Omega \rangle$, $\arrowvert \Psi^{nJPC}_{RPA} \rangle =Q_{nJPC}^{\dagger}(RPA) \arrowvert
\Omega_{RPA} \rangle$,
with $  J $, the total angular momentum, $P$, the parity,
$C$,  the $C$-parity, $n$,  the radial-node quantum number and
\begin{equation}\label{TDAop}
Q^{\dagger}_{nJPC}(TDA) = 
\int \frac{d\bf k}{(2\pi)^3} \Psi ^{nJPC}_{\mu \bar {\mu}}({\bf k})
B^{{\cal C}\dagger}_{\mu}({\bf k}) D^{{\cal C}\dagger}_{\bar {\mu}} (-{\bf k}) 
\end{equation}
\begin{eqnarray} \label{RPA1}
Q_{nJPC}^{\dagger}(RPA)=
\int  \!\!\frac{d{\bf k}}{(2\pi)^3} [X ^{nJPC}_{\mu \bar {\mu}}
B^{{\cal C}\dagger}_{\mu}({\bf k}) D^{{\cal C}\dagger}_{\bar {\mu}} (-{\bf k})
- Y^{nJPC}_{\mu \bar {\mu}}
B^{\cal C}_{\mu}({\bf k}) D^{\cal C}_{\bar {\mu}} (-{\bf k})]   .  
\end{eqnarray}
Note the RPA now involves the RPA vacuum defined by
$Q_{nJPC}(RPA) \arrowvert
\Omega_{RPA} \rangle = 0$. The TDA and coupled RPA equations are given in Refs.~\cite{flsc,flsc3}.
For the vector mesons $\rho$, $J/\psi$, the wavefunctions have been generalized to
included both s and d waves leading  
to four coupled RPA
equations detailed in
Ref.~\cite{flscss}.

Exactly solving the TDA and RPA equations of motion, $H_{CG}\arrowvert \Psi^{nJPC}\rangle = M_{nJPC}  \arrowvert\Psi^{nJPC}\rangle$, yields predictions for the $u/d, s, c$ and $b$ meson spectra in reasonable agreement with observation.  Several results are especially  noteworthy. Except for the pion and light $\eta$, the TDA and RPA spectra were essentially identical. For light mesons, a realistic pion mass ($\approx 150 $ MeV) was obtained along with  Regge trajectories that were also consistent with scattering measurements, 
$J= \alpha(t) =  bt + \alpha(0) \approx .9 t + .5$, $t = M^2_{nJPC}$.  For heavy mesons, the predicted small hyperfine splittings were in excellent agreement with data as summarized in Table 1. Further, the CG model accurately predicted \cite{flscss} the lightest bottomonium state years before its discovery \cite{babar} and much closer than both lattice \cite{lattice} and non-relativistic QCD \cite{nrqcd} results.

\begin{table} [h]
\caption{\label{parameters}
Model comparison of heavy meson predictions with data.} 
\center{

  \begin{tabular}{|c|c|c|c|c|}
    \hline
in MeV &  NRQCD & lattice& Coulomb gauge& data  \\
    \hline
    $\eta_c - J/\Psi$ & 104 & 90 & 125 & 117.7  \\
 $\eta_b - \Upsilon$   &   39$^\dagger$ & 61$^\dagger$ & 70$^\dagger$ & 71.4  \\
$m_{\eta_b} $   &   9421 $\pm$ 11$^\dagger$ & 9409$^\dagger$ &9395  & 9389
\\
  \hline
  \end{tabular}
  \label{table:comparison-u}
\\{\it $^\dagger$ predicted before pseudoscalar bottomonium discovery}
}

\end{table}

It is significant that the CG model is able to simultaneously describe  the small
charmonium/bottomonium  and large $\pi/\rho$ splittings with the same hyperfine
interaction.  This is possible because the RPA meson operator commutes with the chiral charge,
$Q_5  =\int d{\bf x} \Psi^\dagger ({\bf x}) \gamma_5 \Psi ({\bf x})$,
and preserves chiral symmetry yielding a light Goldstone pion.  Conversely, the TDA operator does not commute and the TDA pion for
the same $H_{CG}$ has mass about 500 MeV.  Hence chiral symmetry is predominantly 
responsible for the large $\pi$/$\rho$ splitting, contributing about 400  MeV.  Note that the
model also describes the excited state spin splittings, typically less than 200 MeV,
which are entirely from the hyperfine interaction. These states are not governed by chiral symmetry
and the TDA and RPA results agree to within a few percent.  See Ref. \cite{flscss} for a more detailed discussion of model meson results.

In the gluonic sector, the TDA glueball
wavefunction for two constituent gluons is given by
\begin{eqnarray}
|\Psi^{JPC}_{LS}\rangle= \int{d{\bf
k}\over(2\pi)^3}\Phi^{JPC}_{LS\lambda_1
\lambda_2}({\bf k}) \alpha_{\lambda_1}^{a\dagger}({\bf
k})\alpha_{\lambda_2}^{a\dagger}(-{\bf k})|\Omega\rangle \ .
\label{eq:3.6}\end{eqnarray}
Here $L$ is  the orbital anglular momentum and $S = 0, 1$ or $2$ is the total 
intrinsic gluon spin.
Using this quasiparticle basis the excited glueball spectrum was computed
by diagonalizing the  Hamiltonian in the TDA truncated at the 1p-1h
quasiparticle level.  The RPA spectrum was also calculated and agreed with the TDA to within 1 per cent.  
The predicted $J^{PC} = 0^{\pm +}, 2^{\pm +}$ and $3^{++}$
states agree well with quenched lattice gauge measurements. The predicted $J^{PC} =2^{++}, 4^{++}$ glueballs yielded a Regge trajectories close to the observed pomeron result,
$\alpha_P \approx .25 t + 1$. 
See Ref.~\cite{pom} for further details.
\\

\begin{center}
\center{\Large\em{5.  Oddball and hybrid meson states}}
\\
\end{center}

Predictions for the low lying spectra of $ggg$ glueballs  and $q {\bar q} g$ hybrid mesons  are now presented and discussed.  Since these hadrons consist of 3 constituents, the masses for selected $J^{PC}$ states are computed variationally
\ba
\label{varyeq}
M_{J{PC}}& =& \frac{ \langle \Psi^{JPC} | H_{\rm CG} | \Psi^{JPC} \rangle} {\langle\Psi^{JPC}|\Psi^{JPC}\rangle} \ .
\ea
The variational approximation has been comprehensively tested in two body systems by comparison with exact diagonalization and found to be  accurate to  a few percent.
For $C =  -1$ glueballs (oddballs),  Fock states with at least 3 gluons are necessary and
the   wavefunction  is (${\bf q}_{i = 1,2,3}$  are  $cm$ gluon momenta)
\begin{eqnarray}
\lefteqn{\arrowvert \Psi^{JPC} \rangle = \int d{\bf q}_1 d{\bf q}_2
d{\bf q}_3 
\delta({\bf q}_1 + {\bf q}_2 + {\bf q}_3 )F^{JPC}_{\mu_1 \mu_2 \mu_3}({\bf q}_1,{\bf q}_2,{\bf q}_3) } \nn \\ 
& & 
\hspace{2cm}  
C^{abc}
\alpha^{a\dagger}_{\mu_1}({\bf q}_1)
\alpha^{b\dagger}_{\mu_2}({\bf q}_2) \alpha^{c\dagger}_{\mu_3}({\bf q}_3)
\arrowvert \Omega \rangle \ .
\end{eqnarray}
The
color tensor, $C^{abc}$,   is either 
antisymmetric, $f^{abc}$, for $C = 1$ or symmetric,
$d^{abc}$, for $C = -1$ and
Boson statistics  requires  $C = -1$ oddballs to have a
symmetric space-spin wavefunction.
Using a two-parameter variational radial wavefunction,
the $J^{--}$ oddball states have been calculated. The hyperfine interaction, $H_{qg}$, was suppressed and only the Abelian component of the magnetic fields
was included. 
 The Monte Carlo method with  the
adaptive sampling algorithm VEGAS was used
and numerical
convergence required between  $10^5$ and $10^6$ samples.
In Table 2 results~\cite{lbc} are compared to 
lattice gauge results~\cite{mp,meyerteper} and a Wilson-loop inspired model~\cite{ks}.
The oddball mass sensitivity to both statistical and variational
uncertainties was a few per cent.

\begin{table}[t]
\caption{\label{statetable} Oddball  quantum numbers and  masses
in MeV. Error (Monte Carlo) for  $H_{\rm CG}$  is less than
100 MeV, 
lattice errors are typically 200-300
MeV.}
\center{
\begin{tabular}{|c|cccc|}
\hline
 Model &  $1^{--}$ &  $3^{--}$ & $5^{--}$ & $7^{--}$
\\ \hline
Coulomb gauge~\cite{lbc}&   3950  & 4150 & 5050 & 5900  \\
lattice~\cite{mp}&  3850   &4130 & &  \\
lattice~\cite{meyerteper}& 3100  &4150 & &  \\
Wilson-loop~\cite{ks}& 3490  & 4030 & & \\
\hline
\end{tabular}
}
\end{table}

The predicted oddball Regge trajectories from several
approaches are displayed in Fig. 1.
   Constituent gluon predictions
are represented by boxes, solid  triangles and solid circles and correspond
to   a Wilson-loop inspired potential model~\cite{ks},  a simpler harmonic oscillator
calculation~\cite{lbc}  and
the  Coulomb gauge model~\cite{lbc}, respectively.
Lattice results are depicted by  open
circles~\cite{mp} and diamonds~\cite{meyerteper}.
The   odderon trajectories for the harmonic oscillator and Coulomb gauge models are represented respectively by the solid
lines,  $\alpha_O^{\rm M} = 0.18t + 0.25$ and $\alpha_O^{\rm CG} = 0.23 t - 0.88$,
 while the
$\omega$ trajectory is the
much steeper dashed line.
\begin{figure}[b]
\vspace{-6cm}
\hspace{.5cm}
\includegraphics*[scale=.43]{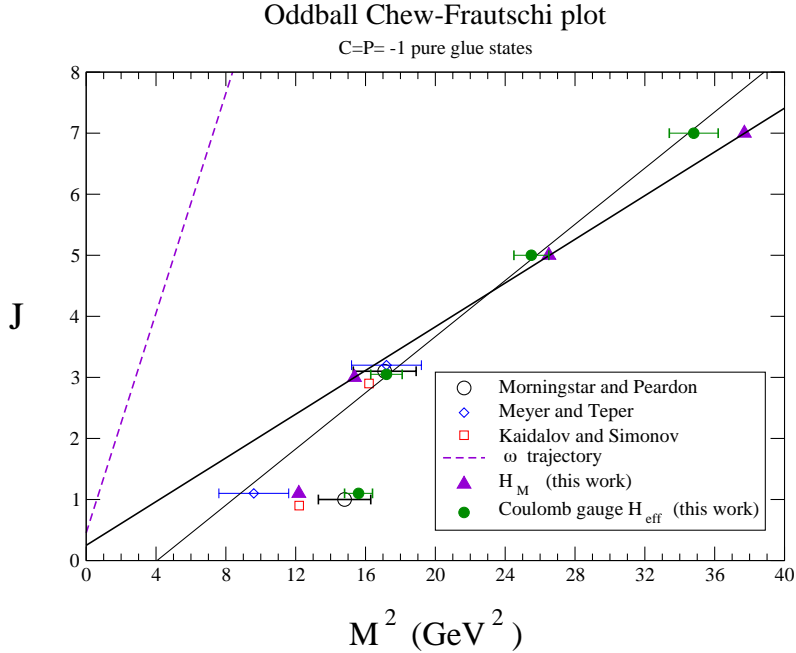}
\vspace{-.8cm}
\caption{Comparison of the $\omega$ meson  Regge trajectory to odderon trajectories from constituent gluon models and lattice.}
\end{figure}

\newpage

Three key results follow which are also supported by a more recent, but simpler constituent gluon model~\cite{Mathieu:2008pb}.  First,     the odderon starts
with the $3^{--}$ state and  not the $1^{--}$ which is
on a daughter trajectory.  Note that there are no lattice
$5^{--}$  glueball predictions which are necessary to confirm this point
and we strongly recommend that
future studies calculate  higher $J^{--}$ states.
Second, all approaches predict the
$3^{--}$ mass is near  4 GeV.
Third, the
predicted   odderon  has slope
similar
to the pomeron but   intercept clearly lower than
the $\omega$ value.  
This provides an explanation for why the odderon has not been observed in total cross section  data which is predominantly governed by the larger $\omega$ intercept. Future searches should therefore focus on differential  cross section measurements, $d \sigma / d t$, at large $t$ where the  odderon trajectory dominates.

Turning to the hybrid meson system,
the non-abelian magnetic field terms are now included.
 For a $q\bar{q}g$ hybrid the color structure  is given by
$SU_c(3)$ algebra,
   $ (3 \otimes \bar{3}) \otimes 8 = ( 8\otimes 8) \oplus (8\otimes 1)
    = 27 \oplus 10 \oplus 10 \oplus 8 \oplus 8
    \oplus 8 \oplus 1$.
Note to obtain an over all color singlet  the
quarks must  be in an octet state, like the gluon,
which produces a repulsive $q \bar{q}$ interaction, confirmed by lattice at short range, that
raises the mass of the hybrid meson. 
Denoting the momenta of the
dressed quark, anti-quark and gluon by ${\bs q}$, $ { \bar{\bs q}}$
and ${\boldsymbol g}$, respectively,
 the
hybrid $cm$ system
wavefunction is
\begin{eqnarray}
\arrowvert \Psi^{JPC} \rangle =  \! \int \! \! \!d{\bf q} d{\bf \bar q}
d{\boldsymbol  g} 
\delta({\bf q} + {\bf \bar q} + {\boldsymbol g} )\Phi^{JPC}_{\lambda \bar \lambda \mu}({\bf q},{\bf \bar q},{\boldsymbol g}) {T^a_{{\cal C}{ \bar {\cal C}} }B^ \dag_{\lambda {\cal C}}({\bf q})
    D^ \dag_{\bar \lambda{ \bar {\cal C}}}}(\bar{\bf q}) \alpha^{a\dag}_{\mu}({ \boldsymbol g}) |\Omega \rangle .
\end{eqnarray}
 
 With two variational parameters, the hybrid mass
was computed  using the Monte Carlo method which required 
about 50 million samples for convergence with error
around
 $\pm$ 50 MeV. 
 The predicted~\cite{glc} low lying  mass spectra for light  hybrid mesons, with both conventional and unconventional (labeled exotica) quantum numbers,
are presented in Fig. 2. 
Note the isospin splitting due to 

\begin{figure} [b]
\vspace{-2.5cm}
\hspace{.3cm}
 \includegraphics[scale=.84]{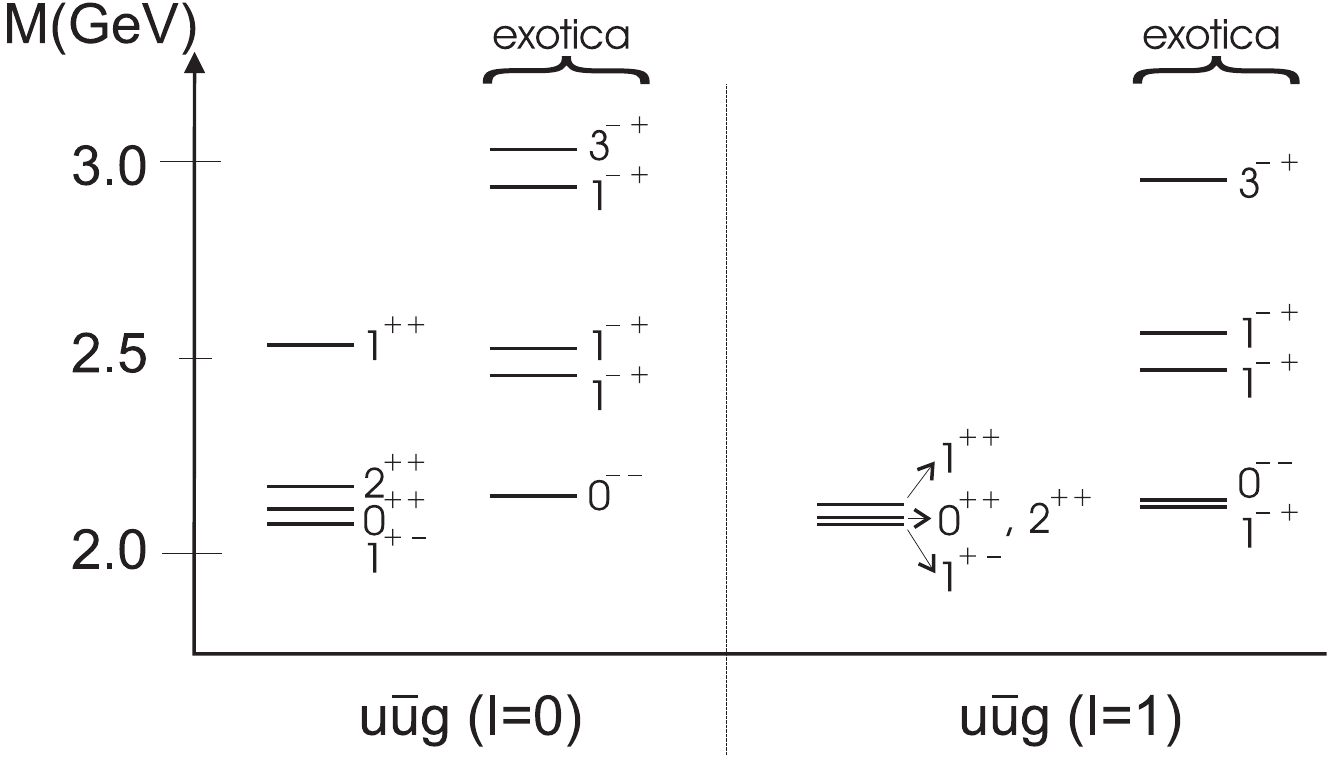}
  \label{fig:resultsUU}
  \caption{Low lying isoscalar and isovector  $u{\bar u}g$ spectra.}
\end{figure}

\newpage

\noindent
 quark annihilation  
(only in the $I =0$ channel) which increases the
 hybrid mass. 
The most significant result is that all hybrid masses, especially the lightest exotic $1^{-+}$ state, are  above 2 GeV.  As summarized in Fig. 3,
this is consistent with lattice  \cite{lattice2,lattice3}  and Flux Tube model \cite{flux}  results
and strongly suggests that the $1^{-+}$ $\pi(1600)$, and more clearly observed $\pi(1400)$, are not hybrid meson states.

\begin{figure} [h]
 \vspace{-3.3cm}
      \hspace{-2 cm}
        \includegraphics[scale=.59]{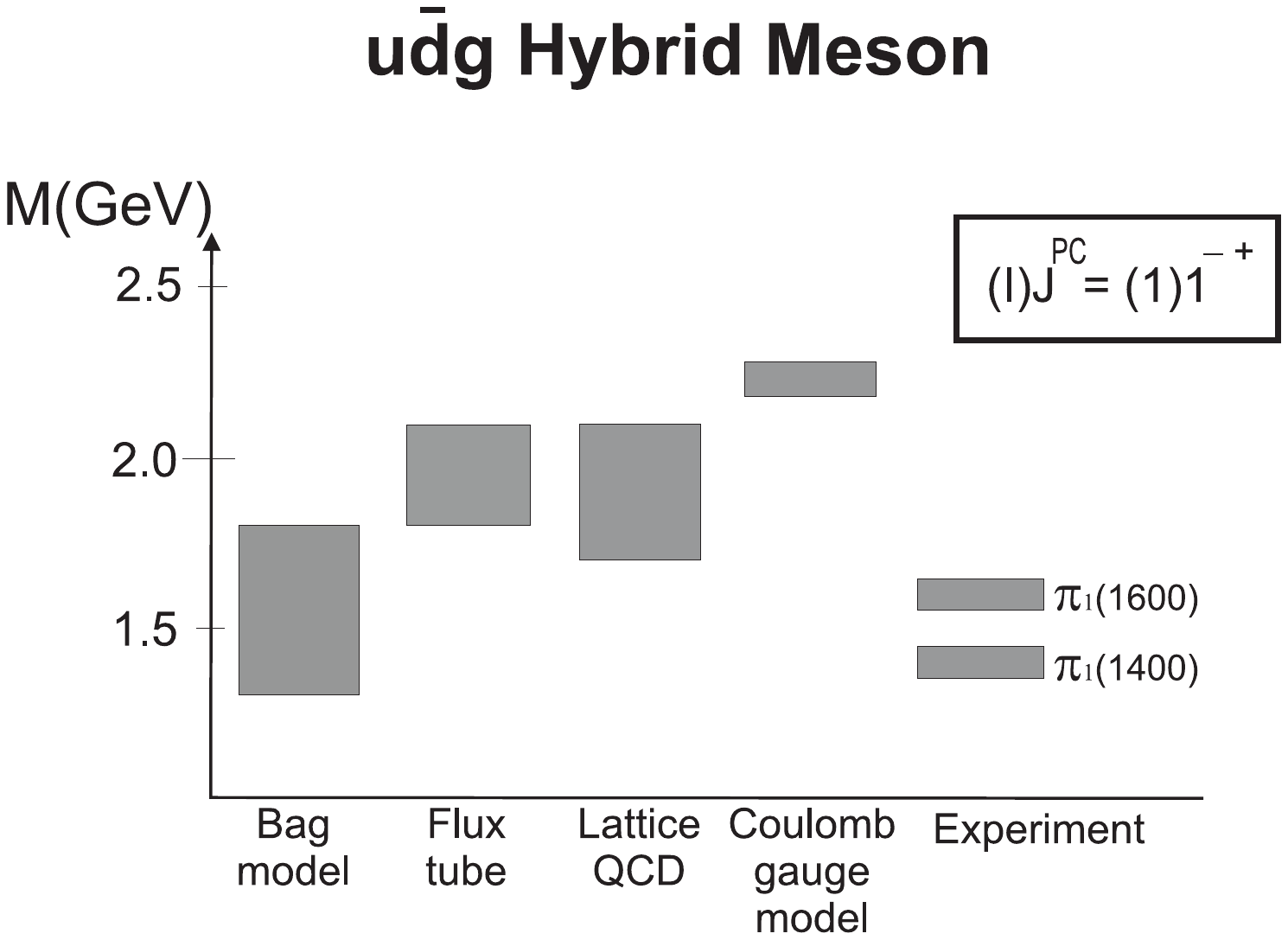}
         \vspace{-3.6cm}
        \caption{ Confrontation of different hybrid meson models with data.}
\end{figure}

\begin{figure} [b]
        \vspace{-3cm}
\hspace{.8cm}
   \includegraphics[scale=.83]{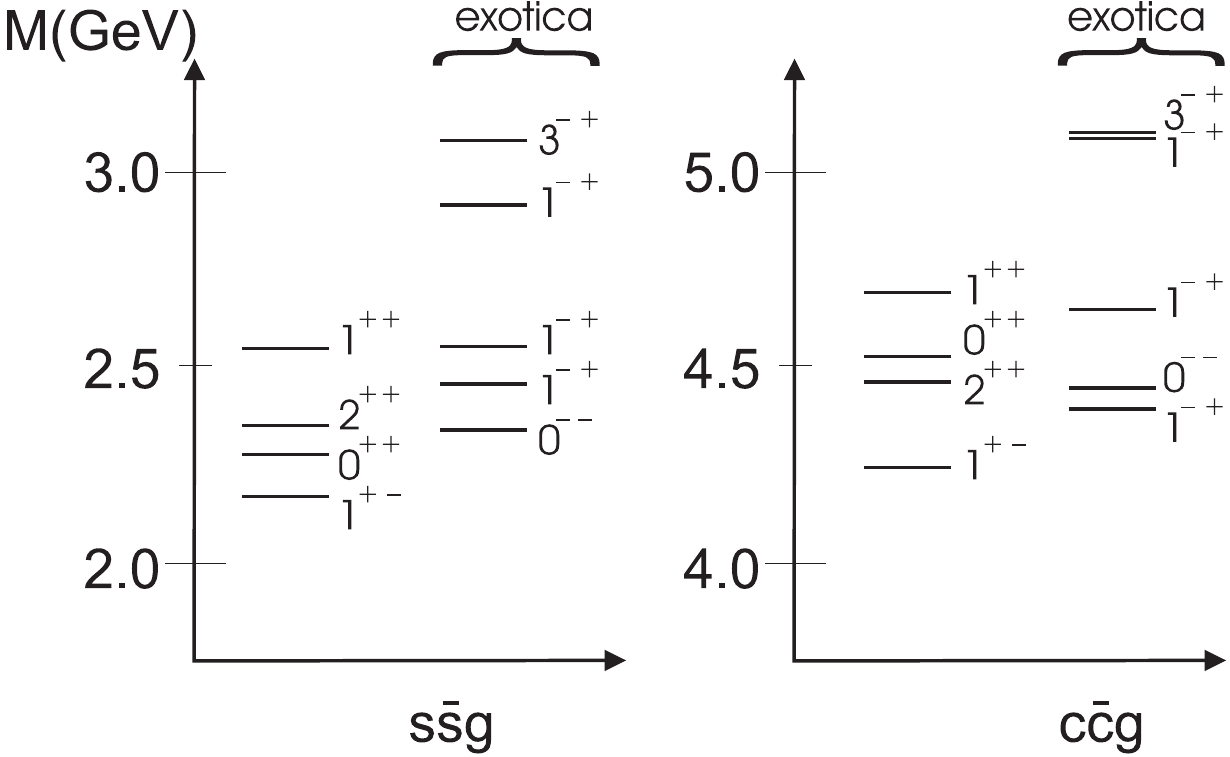}
    \label{fig:resultsSS-CC}
      \caption{Selected low lying $s{\bar s}g$ and $c {\bar c}g$ spectra.}
\end{figure}

 The strange $s \bar s g$ and charmed $c \bar c g$ hybrid spectra are illustrated in Fig. 4.
Due to the hyperfine interaction, the $c \bar c g$ spectra has a slightly different level ordering than the $u \bar u g$ and $s \bar s g$ results.
The  $s \bar s g$ and  $c \bar c g$ exotic $1^{-+}$ states are also in reasonable agreement with
both lattice and Flux Tube results.
\vspace{.75cm}






\newpage

\begin{center}
\center{\Large\em{6.  Tetraquark states}}
\\
\end{center}

As illustrated in Fig. 5,
the $SU_c(3)$ color algebra for four quarks produces 81 color states,
$3\otimes\bar3\otimes3\otimes\bar3=27\oplus10\oplus
\bar{10}\oplus8\oplus8\oplus8\oplus8\oplus1\oplus1$. Two are color singlets that
can be obtained in four different ways depending on the
intermediate color coupling:  singlet scheme, non-exotic meson-meson molecule, and
three exotic atoms, one
octet  and two diquark schemes involving the triplet  and the sextet representations.
\begin{figure} [h]
\vspace{-2.5cm}
 \includegraphics[width=1\textwidth]{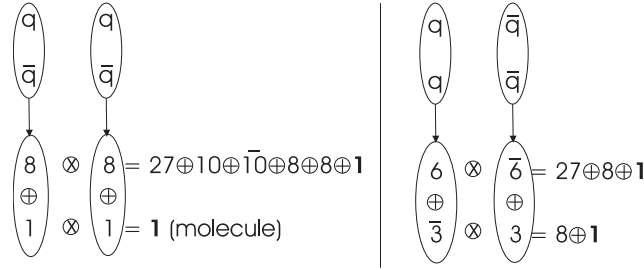}
   \vspace{-3.5cm}
  \label{fig:resultsSS-CC}
 \caption{Color singlets from four different representations.}
\end{figure}
\noindent
The two color singlets, $\delta_{c_1c_2}\delta_{c_3c_4}$ and $\delta_{c_1c_4}\delta_{c_3c_2}$, are linearly independent and form a  color space  as depicted in Fig. 6 with the first  along  the horizontal axis and the second,  in the limit of large $N_c$,
 vertical. 
For physical $N_c=3$ they are not orthogonal, as 
the second is now rotated with respect to the first held fixed. 
However they still span the entire color space so that any of the four schemes  can be represented as a linear combination of the two singlets. This means that the color degree of freedom does not  forbid a tetraquark transition into two mesons.

\begin{figure} [b]
\vspace{-3.5cm}
\hspace{3cm}
 \includegraphics[scale=.6]{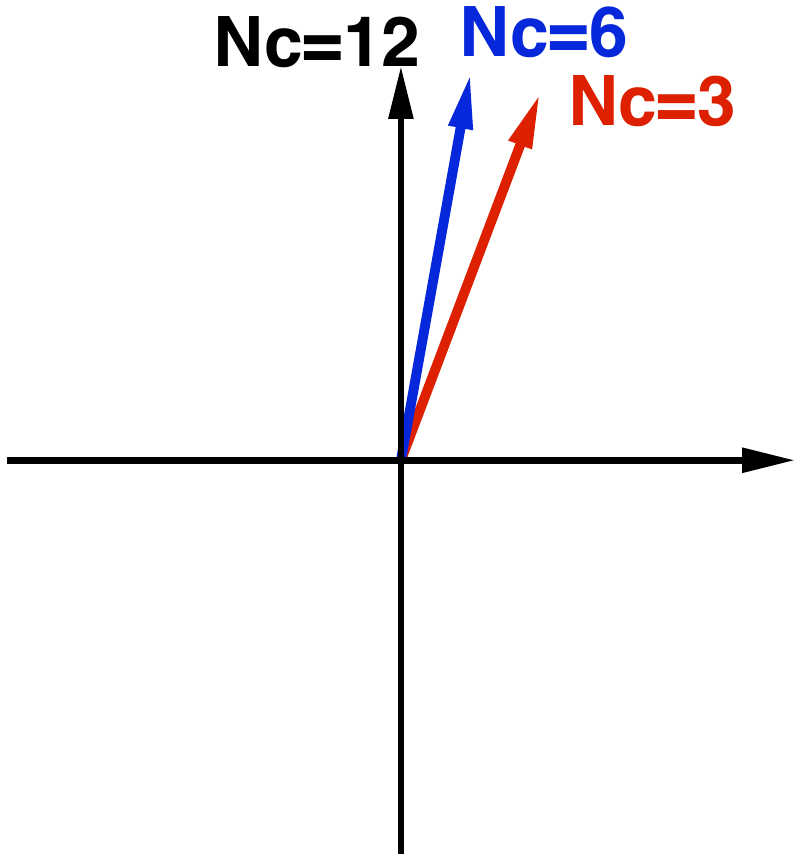}
   \vspace{0cm}
  \label{fig:resultsSS-CC}
 \caption{Color space.}
\end{figure}

\newpage

Denoting
the momenta of 
the quarks by ${\bf q_1}$, ${\bf q_3}$, and
those of the anti-quarks by ${\bf q_2}$, ${\bf q_4}$,
the $cm$ tetraquark wavefunction is
 \ba
\lefteqn{  \hspace{-1.5cm}  |\Psi^{JPC}\rangle= \int \!\!
    {d\boldsymbol{q}_1}{d\boldsymbol{q}_2}
    {d\boldsymbol{q}_3} {d\boldsymbol{q}_4}
    \delta({\bf q}_1 + {\bf q}_2 + {\bf q}_3 + {\bf q}_4 ) \Phi^{JPC}_{\lambda_1 \lambda_2
    \lambda_3 \lambda_4}({\boldsymbol{q}_1,\boldsymbol{q}_2,\boldsymbol{q}_3, \boldsymbol{q}_4})  } \nonumber  \\
    & &
  R^{{\cal C}_1{\cal C}_2}_{{\cal C}_3{\cal C}_4}   B^{\dag}_{\lambda_1{\cal C}_1}({\boldsymbol{q}_1})
   { D^{\dag}_{\lambda_2{\cal C}_2}(\boldsymbol{q}_2)}
   { B^{\dag}_{\lambda_3{\cal C}_3}(\boldsymbol{q}_3)}
    {D^{\dag}_{\lambda_4{\cal C}_4}(\boldsymbol{q}_4)}|\Omega \rangle \ , 
  \ea
  where the color elements, $R^{{\cal C}_1{\cal C}_2}_{{\cal C}_3{\cal C}_4}$,
depend on the specific color scheme chosen.
Contributions to the Hamiltonian expectation value involve 
4 self-energy, 6 scattering, 4
annihilation and 70 exchange terms each of which can be reduced to 12 dimensional integrals that are evaluated in momentum space. Again, these were computed~\cite{gwcl} by performing Monte Carlo calculations (typically 50 million samples) and the hyperfine interaction was not included.
The  meson-meson molecule yields the lightest mass state for a given $J^{PC}$ which is due to cancellation of certain interactions by color factors in the singlet-singlet molecular representation and also the presence of repulsive forces in the other,
more exotic,  color schemes.
 The  ground state is a
non-exotic  $1^{++}$ 
with  mass around 1.2 GeV. The remaining low lying spectra for states having both conventional 
and unconventional (loosely labled exotica but not exotic) quantum numbers in the molecular singlet  color representation is displayed in Fig. 7. 
As in the hybrid calculation, there are isospin splitting contributions, up to several hundred MeV, but only in the octet scheme (not shown)
from quark annihilation interactions ($q \bar{q} \rightarrow
g \rightarrow  q \bar{q}$) in the $I_{q \bar{q}} = 0$ channel.  
The annihilation interaction is repulsive, yielding octet states with $I =2$  lower than the $I = 1$ which are lower than the $I = 0$.  The meson-meson states are all  isospin degenerate producing several molecular
tetraquark states with the same $J^{PC}$ in the 1 to 2 GeV 
\begin{figure} [b]
\vspace{-6.78cm}
\hspace{-3.7cm}
\includegraphics[scale=.68]{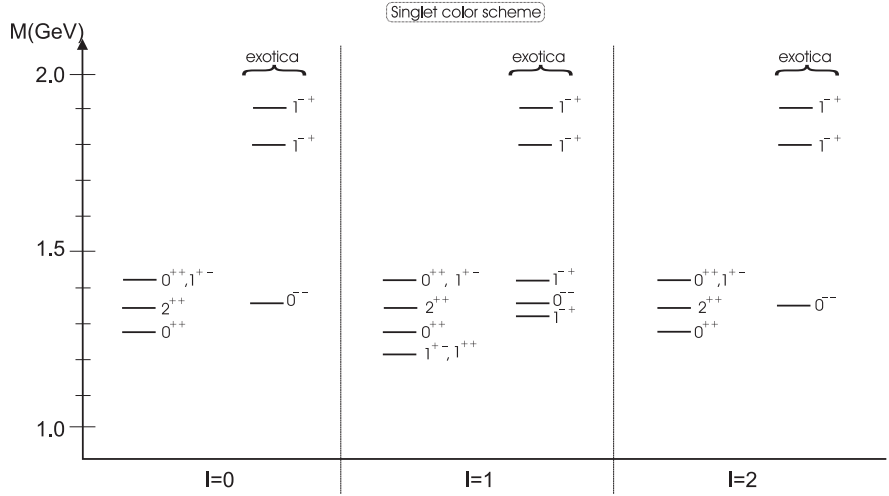}  
  \vspace{-4.4cm}
    \label{fig:resultsSS-CC}
     \caption{Non-exotic tetraquark  color singlet (molecule) spectra.}
\end{figure} 

\newpage
\noindent
region.  The 
$I = $ 1 and 2 states can be observed in different electric charge channels (different $I_z$) at about the same energy, which is
a useful experimental signature.   
Most important,  the predicted
lightest $1^{-+}$ is  at 1.32   GeV,  near the observed $\pi(1400)$, suggesting this state has a non-exotic meson-meson resonance molecular structure. Indeed, the predicted tetraquark mass for all exotic $1^{-+}$ states in the octet color configurations is above 2 GeV.  This is consistent with  the
 model predictions, discussed in the previous section, 
for exotic hybrid meson  $1^{-+}$ states also lying above 2 GeV due to repulsive octet color  quark
interactions.  Note that $C$ parity forbids exotic diquark ($[3 \otimes  3] \otimes [\bar 3 \otimes \bar 3]$)   states in the $1^{-+}$ channel.
 The other $J^{PC}$ states  in both the triplet  and the sextet diquark color representations
 also  have computed masses (not shown) heavier than in the singlet representation and  comparable to the octet
scheme results. 
Since there are  two linearly independent color configurations, in lieu of a color mixing calculation that would order the  eigenvalues for a given basis, the variational procedure 
requires accepting as the best approximation to the physical states the molecular scheme as the lightest along with the  heavier    linear combination that is orthogonal.

Finally, the important issue of meson and tetraquark mixing is addressed~\cite{wcg}
for the $J^{PC}=0^{\pm +}$ and $1^{--}$ states. 
Using the notation, $|q \bar{q}>$ and $|q \bar{q} q \bar{q}>$ for $|\Psi^{JPC}>$, the  mixed state is  given by
$
|J^{PC}\rangle =
a|n\bar{n}\rangle+b|s\bar{s}\rangle+c|n\bar{n}n\bar{n}\rangle+d|n\bar{n}s\bar{s}\rangle,
$
where $n\bar{n}=\frac1{\sqrt{2}}(u\bar{u}+d\bar{d})$. The state
$| s\bar{s}s\bar{s}\rangle$ is not included since its
mass is much higher. The coefficients $a,b,c$
and $d$ are determined by diagonalizing the Hamiltonian matrix.
The off-diagonal mixing element, $M = {\langle q\bar{q}|H_C^{\rm CG}|q\bar{q}q\bar{q}\rangle}$, only involves $H_C^{\rm CG}$ which connects meson and tetraquark states. 
Of the six off-diagonal matrix elements  two,
$\langle s\bar{s}|H_C^{\rm CG}|n\bar{n}\rangle$ and $\langle s\bar{s}|H_C^{\rm CG}|n\bar{n}n\bar{n}\rangle$,
vanish and one, $\langle n \bar{n}  n\bar{n}|H^{\rm CG}_C|n\bar{n} s \bar{s}\rangle$, is computed very small.
The remaining   three
elements  are, $\langle n\bar{n}|H^{\rm CG}_C|n\bar{n}n\bar{n}\rangle$,
$\langle n\bar{n}|H_C^{\rm CG}|n\bar{n}s\bar{s}\rangle$ and $\langle
s\bar{s}|H_C^{\rm CG}|n\bar{n}s\bar{s}\rangle$. Because
of color factors,  nonzero mixing only exists   for
$q \bar{q}$ annihilation   between  different singlet
$q \bar{q}$ clusters. 
 There are two contributions to each mixing matrix element.  
 One is
\begin{eqnarray}
&& M_1 = \frac12 \int\!\!\! \; d{\boldsymbol
q}_1^{} d{\boldsymbol q}_2^{} d{\boldsymbol q}_3^{}V(k)
{\cal U}_{\lambda_1}^\dag({\boldsymbol q}_1^{})
{\cal U}_{\lambda'_1}(-{\boldsymbol q}_4) \\ \nonumber &&
{\cal U}_{\lambda_3^{}}^\dag({\boldsymbol q}_3^{})
{\cal V}_{\lambda_2^{}}({\boldsymbol q}_2^{})\Phi_{\lambda_1^{}
\lambda_2^{} \lambda_3^{}\lambda_4^{}}^{JPC\dag}({\boldsymbol
q}_1^{},{\boldsymbol q}_2,{\boldsymbol q}_3^{})
\Phi_{\lambda'_1\lambda_4}^{JPC}(-2{\boldsymbol
q}_4),
\end{eqnarray}
with
$V(k)$ the
confining potential,
${\boldsymbol q}_4= -{\boldsymbol q}_1 - {\boldsymbol k}$,  ${\boldsymbol
k}={\boldsymbol q}_2+{\boldsymbol q}_3$ and 
\begin{eqnarray}
{\cal U}_\lambda &= &\frac1{\sqrt{2}}\left (
\begin{array}{lcr}
\sqrt{1+s_q} \\
\sqrt{1-s_q} \; \;  {\boldsymbol \sigma}\cdot  {\boldsymbol {\hat q}} \\
\end{array}
\right )\chi_\lambda \\
{\cal V}_\lambda &=& \frac1{\sqrt{2}}\left (
\begin{array}{lcr}
-\sqrt{1-s_q}  \; \; {\boldsymbol \sigma}\cdot  {\boldsymbol {\hat q}} \\
\sqrt{1+s_q} \; \;  \\
\end{array}
\right )\chi_\lambda  \ ,
\end{eqnarray}
are dressed, BCS spinors.
Again $s_q = sin \, \phi_q$ is obtained from the gap equation solution.
The   other contribution has the form
\begin{eqnarray}
&& M_2 = \frac12 \int\!\!\! \; d{\boldsymbol
q}_1 d{\boldsymbol q}_2 d{\boldsymbol q}_3 V(k)
{\cal V}_{\lambda_4}^\dag({\boldsymbol q}_4)
{\cal V}_{\lambda'_4}({-\boldsymbol q}_1) \\ \nonumber &&
{\cal U}_{\lambda_3}^\dag({\boldsymbol q}_3)
{\cal V}_{\lambda_2}({\boldsymbol q}_2)\Phi_{\lambda_1
\lambda_2 \lambda_3\lambda_4}^{JPC\dag}({\boldsymbol
q}_1,{\boldsymbol q}_2,{\boldsymbol q}_3)
\Phi_{\lambda_1\lambda'_4}^{JPC}(2{\boldsymbol q}_1) \ .
\end{eqnarray}
 Because  new model masses are computed, the unmixed variational basis states
need not be ones producing a minimal, unmixed mass.  This allows  adjusting one of the two
variational parameters, denoted by $\gamma$, to provide an optimal variational mixing prediction. 
For $0^{++}$ states,
the  mixing
term vanishes for $\gamma =$ 0  and then  increases with
increasing $\gamma$.
For the optimum variational value, $\gamma=0.2$, the  matrix elements are
$\langle s\bar{s}|H_C^{\rm CG}|n\bar{n}s\bar{s}\rangle =365$ MeV,
$\langle n\bar{n}|H_C^{\rm CG}|n\bar{n}n\bar{n}\rangle =166$ MeV and
$\langle n\bar{n}|H_C^{\rm CG}|n\bar{n}s\bar{s}\rangle =45$ MeV.
With the calculated matrix elements and   unmixed meson and tetraquark masses,  the complete Hamiltonian matrix was diagonalized to obtain the expansion  coefficients and masses
for the corresponding eigenstates illustrated in Fig. 8.  Mixing  
clearly provides an
improved description  for the $f_0$ spectrum
\begin{figure}[t]
\vspace{-1cm}
\hspace{2cm}
\includegraphics[scale=.5]{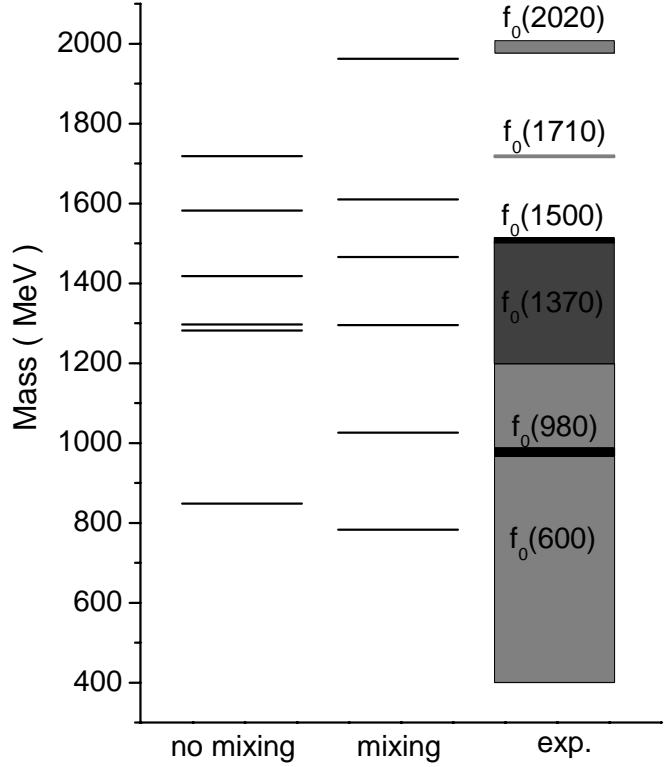}
\vspace{-1.5cm}
\caption{Unmixed and mixed $f_0$ spectrum compared to data.}
\end{figure}
as the
$\sigma$ meson mass is lowered from 848 MeV to 776 MeV,  the strange
scalar meson mass  decreases from 1297 MeV to 1006 MeV,
closer to the observed value of 980 MeV, and the other
$f_0$ states are now also in better agreement with data.  
Including chiral symmetry corrections, which are omitted by the TDA variational basis states,
will further lower the $\sigma$ prediction, closer to the now accepted value  of 450 MeV
\cite{GarciaMartin:2009aq}.
The mixing coefficients also provide new structure insight, predicting  the $\sigma / f_0(600)$ is predominantly a mixture 
of $n\bar{n}$ and $n\bar{n}n\bar{n}$ states while the $f_0(980)$
 is mainly  $s\bar{s}$ and
$n\bar{n}s\bar{s}$ states.
This is consistent with
the growing consensus  that the $\sigma / f_0(600)$ state
is a $\pi \pi$ resonance (pole in the $\pi \pi$ scattering amplitude) with a
 molecular tetraquark nature.
 Future mixing calculations will include both scalar glueballs and hybrid mesons which
 should further improve describing the high lying $f_0$ spectrum, anticipated to have a few newly discovered states, and also aid identification of gluonic states.  

For $0^{-+}$ states, 
the
value $\gamma = 0.5$ yields
reasonable  $\eta$ and $\eta{'}$ masses with mixing 
elements  $\langle n\bar{n}|H_C^{\rm
CG}|n\bar{n}n\bar{n}\rangle= 219$ MeV, $\langle n\bar{n}|H_C^{\rm
CG}|n\bar{n}s\bar{s}\rangle= 157$ MeV and $\langle s\bar{s}|H_C^{\rm
CG}|n\bar{n}s\bar{s}\rangle=138$ MeV. The unmixed $\eta$, $\eta'$ masses
changed from 610 MeV, 1002 MeV to 531 MeV, 970 MeV, respectively, both
closer to the observed values of 547.51 MeV  and 957.78 MeV.

A novel mixing result was obtained for the $1^{--}$ states.
Again the mixing matrix elements were   0 for $\gamma = 0$ but,
and very interesting,
 also  essentially 0 for all values of 
$\gamma$.  The Coulomb gauge model therefore predicts minimal flavor  mixing 
for vector mesons which 
agrees with the
known,  predominantly ideal, $\omega/\phi$ mixing.  Related, the model still provides a good vector meson spectrum description
since the unmixed $n \bar{n}$ and $s \bar{s}$  states were already  in  agreement \cite{flscss} with observation.
\\

\begin{center}
\center{\Large\em{7.  Summary}}
\\
\end{center}


Concluding, the Coulomb gauge model  provides a comprehensive, unified
quark-gluon framework for realistically describing the vacuum and  meson spectrum 
and also agrees with
glueball and hybrid meson predictions from alternative approaches.
The model is sufficiently robust, as evidenced by accurately predicting the $\eta_b$ mass
and attending hyperfine splitting,
to guide experimentalists in future particle searches, especially states with explicit gluonic degrees of freedom.  Further, this approach retains the attractive wavefunction picture, not available through lattice QCD, which provides deeper hadronic insight  that should be helpful
in understanding states with unconventional quantum numbers like the observed $\pi_1$ which appears to be a meson-meson resonance.   
Finally, the approach is  amendable to further refinements through improved confining interactions and extended model spaces involving additional quasiparticle Fock states.

\begin{center}
{\center{\small{\em Acknowledgments}}}
\end{center}
\indent
Thanks to the NAPP2010  
organizers, especially B. Vlahovic and I. Supek.
I. General and P. Wang are gratefully acknowledged. 
Support is from grants U. S. DOE  DE-FG02-03ER41260,
FPA 2008-00592, FIS2008-01323 plus 227431, Hadron-Physics2 (EU) and PR34-1856-BSCH, UCM-BSCH GR58/08, 910309 and PR34/07-15875.
\newpage

%

\end{document}